\documentclass[%
 reprint,
 amsmath,amssymb,
 aps,
]{revtex4-1}

\usepackage{graphicx}
\usepackage{dcolumn}
\usepackage{bm}
\usepackage{tensor}

\begin{document}

\title{Relaxation Principle in $(n+1)$ Dimensions}

\author{Choong Min Um}
\affiliation{Department of Physics, University of Washington, Seattle, WA 98195, USA}

\date{\today}

\begin{abstract}
Why do we live in a $(3+1)$ dimensional universe? In this paper, I review the ``relaxation principle" first introduced in \cite{karch} and generalize its ideas to an arbitrary $(n+1)$ dimensions. This is done by referring to the Friedman equation and the scaling solution to derive the energy densities of the non-interacting and interacting branes, and then formulating their dimensionalities. I also demonstrate that the largest interacting $d$-brane in $n$ spatial dimensions is always the $(n-2)$-brane, and that such dimensionality constraint ``relaxed" our universe from nine dimensions to three dimensions.
\end{abstract}

\maketitle

\section{Introduction}
One ought to wonder why our universe is $(3+1)$ dimensional because there is no law of nature that is unique to $(3+1)$ dimensions of spacetime. In \cite{brandenberger}, this question has been addressed from a string theory perspective. Instead of the ordinary ``spontaneous compactification" idea, Brandenberger and Vafa suggested a new scenario in which all the dimensions started out small and compact, and only three spatial dimensions ``decompactified" to grow large. They argued that the winding modes of a string ``try" to prevent expansion because increasing the volume increases the string winding energy. Assuming periodic boundary conditions, a string spans a $2$ dimensional worldsheet, and therefore two strings intersect only in $n\le(3+1)$ dimensions, after which branes can unwind and expand. They claimed that this happened to be our observable universe. While this idea was interesting and aligning well with the Big Bang theory, it also came with some setbacks, as pointed out in \cite{karch}. Namely, 1) it relied on the poorly understood Planck scale dynamics; 2) it required a resolution of the moduli problem; 3) it relied on simple toroidal compactification; and 4) it relied on strings as having the only fundamental degrees of freedom, whereas now it has become clear that branes do, too. In \cite{alexander}, Alexander et al. incorporated the role of branes in the decompactification idea.

On the other hand, \cite{durrer} argued that at very early times the space was already large, filled with $d$-branes and $\tilde{d}$-antibranes. The authors stated that because $4+4<10$, the worldvolumes of 3-branes would not intersect (thus, non-interacting branes) whereas the worldvolumes of higher dimensional branes would (interacting branes). Generally, when a $d$-brane intersects with another $d$-brane on a $(d-1)$ hypersurface, one side of the first brane reconnects with the other side of the second brane until their winding mode is reduced and the branes evaporate, which ultimately increases the entropy of the universe. Their idea was also very interesting, but \cite{karch} again pointed out a flaw in their evaporation mechanism that the intersected branes would generally merge to form new $(p,q)$-type branes, not just evaporate.

Finally, we have the ``relaxation principle" from \cite{karch}. The authors stated that the universe was initially filled with branes of all sizes but naturally selected among possible vacua the configuration with the biggest filling fraction (or energy density), which in our $(9+1)$ dimensional universe turned out to be the 3-branes and 7-branes. Other branes simply diluted, after dissipating their energy at the maximal rate allowed by causality. They also argued that the energy-momentum of the branes drived the cosmological evolution and the brane dynamics determined the number of dimensions that we see. This paper builds upon their ideas and see what can be said about the relaxation principle in $n$ dimensional universe in general.

\section{The FRW Universe}
We believe that our observable $(3+1)$ dimensional universe is spatially homogeneous and isotropic, and evolving in time, and such universe is described by the Friedmann-Robertson-Walker (FRW) metric:
\begin{equation}\label{reio}
ds^2=-dt^2+a^2(t)\left(\frac{dr^2}{1-\kappa r^2}+r^2d\Omega ^2\right)
\end{equation}
where $a(t)=R(t)/R_0$ is the dimensionless scale factor and $\kappa=k/R_0^2$ is the curvature parameter. If we choose to model matter and energy by a perfect fluid, the energy-momentum tensor becomes $T\indices{^\mu_\nu}=diag(-\rho, \vec{p})$ and plugging this into the Einstein's equation yields the Friedman equation. Generalizing into $n$ spatial dimensions, we get
\begin{equation}
\left(\frac{\dot{a}}{a}\right)^2=\frac{16 \pi G}{n(n-1)}\rho-\frac{\kappa}{a^2}
\end{equation}
where $\rho$ is the energy density, $p$ is the isotropic pressure and $\dot{a}/a$ is the rate of expansion, or the Hubble parameter. Since perfect fluids obey the simple equation of state $p=w\rho$, the conservation of energy equation dictates
\begin{equation}
\frac{\dot{\rho}}{\rho}=-n(1+w)\frac{\dot{a}}{a}
\end{equation}
which can be solved (given that $w$ is a constant) to give the time dependence of $a$ and $\rho$:
\begin{equation}
\rho \sim a^{-n(1+w)}\label{rho}
\end{equation}
\begin{equation}
t \sim a^{n(1+w)/2}\label{t}
\end{equation}
Some of the choices of $w$ are 1) $w=0$ for matter (dust), which is any set of collisionless particles such as stars and galaxies whose pressure is negligible in comparison with the energy density; 2) $w=1/n$ for radiation, which is either electromagnetic radiation or massive particles moving at relativistic speeds; and 3) $w=-1$ for both vacuum and strings in one dimension.

\section{Strings and Branes}
A string in one dimension has $p=-\rho$, so a string in three dimensions in random direction gives an average $p=-\rho/3$. We can generalize this argument to state that a $d$-brane with a $(d+1)$ dimensional worldvolume in $n$ spatial dimensions has $w=-d/n$. From this, it immediately follows that the energy density of a non-interacting gas (fluid) of $d$-branes goes as
\begin{equation}
\boxed{\rho^{ni}_d \sim a^{-n(1-d^{ni}/n)}=a^{d^{ni}-n}}\label{ni}
\end{equation}
To see how the energy density of an interacting gas behaves, we make use of the horizon scale in which the network of branes at any time appears the same when scaled to the Hubble radius. For example, a scaling solution for a string implies that the length of the string evolves proportionally to $t$, i.e., $l_s(t) \sim t$. Similarly, the total volume of a $d$-brane is $t^d$ and the horizon volume is $t^n$, hence the energy density of an interacting gas of $d$-branes must be
\begin{equation}
\boxed{\rho^i_d \sim t^{d^i-n}}\label{i}
\end{equation}
According to \cite{alexander}, whether or not two branes intersect depends solely on their dimensionality. In $(n+1)$ dimensions, generic $d$-branes with: (i) $2d \geq n$ intersect at all times over a $(2d-n)$-brane; (ii) $2d=n-1$ intersect over a $(-1)$-brane; and (iii) $2d \leq n-2$ do not find each other. Strings in $n=3$, for example, would go into the (ii) category. For $n=9$, the critical dimension of superstring theory, it is clear that only the branes with $d \geq 4$ will intersect. The key idea here is that if there is no brane intersection, the higher dimensional branes with higher energy densities would dilute more slowly than the lower dimensional branes. However, the intersections affect the dilution so that higher and lower dimensional branes can compete. When branes intersect, they dilute faster by dissipating energy through spawning loops of closed strings (or balls, in higher dimensions), which then decay by emitting gravitational waves \cite{polchinski}. In odd $n$ dimensions, only the branes with $d \leq \frac{n-3}{2}$ will not intersect. This is because $d \leq \frac{n-2}{2}$ is not an integer when $n$ is odd. For an odd $n$, the largest integer less than $\frac{n-2}{2}$ is $\frac{n-3}{2}$.

\section{The Dominant Branes}
In \cite{karch}, Karch and Randall showed that under FRW evolution, our $(9+1)$ dimensional universe filled with equal numbers of branes and antibranes naturally selected the 3-branes and 7-branes to be dominant through its cosmological evolution. Among the non-interacting branes, 3-brane is the dominant brane because it is the highest dimensional brane that obeys $4+4<10$. Among the interacting branes, 9-branes generically annihilate since they overlap completely at all times, while 8-branes disappear because either 1) even dimensional branes do not exist according to type II string theory; or 2) in flat spacetime static 8-brane/anti-brane configuration cannot have branes far away from each other or else the spacetime breaks down. This leaves us with the 7-brane as the dominant interacting brane. They argued that the energy-momentum of the branes dominated the cosmological evolution, and therefore branes with the biggest filling fraction in the universe turned out to be the ones we live in. The next step is to apply their ideas to an arbitrary $(n+1)$ dimensional universe, and see what can be said about the relaxation principle in general.

In odd spatial dimensions, branes dilute as (\ref{ni}) for $d \leq \frac{n-3}{2}$, and as (\ref{i}) for higher $d$. Among the branes with $d \leq \frac{n-3}{2}$, it is obvious that 
\begin{equation}
\boxed{d^{ni}=\frac{n-3}{2}}\label{d^ni}
\end{equation}
will dilute most slowly because it is the highest dimensional brane with the highest energy density. In the case of $n=9$, we obtain $d^{ni}=3$.

For any $w<w_{crit}=-\frac{n-2}{n}$, (\ref{t}) states that $t$ grows slower than $a$. If we take the $\left(\frac{n-3}{2}\right)$-brane energy density to dominate, then
\begin{equation}
w=-\frac{d}{n}=-\frac{n-3}{2n}
\end{equation}
The resultng time-dependence of the scale factor is
\begin{equation}
t \sim a^{n(w+1)/2}=a^{(n+3)/4}\label{ta}
\end{equation}
Next, we determine which of the interacting branes can compete with the non-interacting $\left(\frac{n-3}{2}\right)$-brane. The two branes can compete only when $\rho^{ni}_d \sim \rho^i_d$. or
\begin{equation}
a^{d^{ni}-n} \sim t^{d^i-n}
\end{equation}
Using (\ref{ta}) to replace $t$ with $a$, this is equivalent to
\begin{equation}
a^{d^{ni}-n} \sim a^{(d^i-n)\left(\frac{n+3}{4}\right)}
\end{equation}
Equating and solving for $d^i$, we get
\begin{equation}
d^i=\frac{4(d^{ni}-n)}{n+3}+n
\end{equation}
Plugging in $d^{ni}=\frac{n-3}{2}$ gives
\begin{equation}
\boxed{d^i=n-2}\label{d^i}
\end{equation}
This equation states that the dominant interacting $d$-brane in $n$ dimensions is always $(n-2)$-brane. The constraint $\rho^{ni}_d \sim \rho^i_d$ ensures that (\ref{d^ni}) and (\ref{d^i}) compete each other in terms of their rates of dilution. In the familiar case of $n=9$, we obtain $d^{ni}=3$ and $d^i=7$ in accordance with \cite{karch}. In even $n$ dimensions, we have $d^{ni}=\frac{n-2}{2}$, $w=-\frac{n-2}{2n}$, and $t \sim a^{(n+2)/4}$, all of which are different values than what we obtained above for odd $n$ dimensions, but $d^i=n-2$ remains invariant.

Our observable universe appears to be $(3+1)$ dimensional, which leaves us with two candidates for the overall dimensionality of our universe: $n=9$ or $n=5$. $n=9$ was discussed above, but what about $n=5$? $n=5$ gives rise to $d^{ni}=1$ and $d^i=3$, which makes our 3-brane an interacting one. Is this a possible scenario? If we can detect and identify some decay mechanisms, such as gravitational waves, perhaps. Is there any indication that nature favors $n=9$ instead of $n=5$ outside the context of string theory? These are some of the questions that I hope to answer in near future.

\section{Conclusion}
In this paper, I reviewed the relaxation principle and generalized its ideas to $n$ spatial dimensions. By using the solutions of Friedman equation and referring to the scaling solution, I derived the energy densities of the non-interacting and interacting branes, from which I formulated the dimensionalities of the dominant branes. I also demonstrated that the largest interacting $d$-brane in $n$ dimensions is always the $(n-2)$-brane, and that for $n=9$ the dominant branes are 3-branes and 7-branes. Whether or not $n=5$ is a feasible number of dimensions in the universe will be studied in future publications.

\section{Acknowledgements}
I would like to thank Professor A. Karch for guidance and discussions.

\end{document}